\documentclass[aps,amssymb,amsfonts,twocolumn,prl,floatfix,showpacs,superscriptaddress,10pt]{revtex4-1}

\usepackage{graphicx}
\usepackage{bm}
\usepackage{amssymb}
\usepackage{amsmath}
\usepackage{epsfig}
\usepackage[T1]{fontenc}
\usepackage{appendix}
\usepackage{enumerate}
\usepackage{aeguill}
\usepackage{subfigure}
\usepackage[utf8]{inputenc}

\begin{document}

\title{Spin-Glass Model Governs Laser Multiple Filamentation}

\author{W. Ettoumi}
\email{wahb.ettoumi@unige.ch}
\affiliation{Universit\'e de Gen\`eve, GAP-Biophotonics, Chemin de Pinchat 22, CH-1211 Geneva 4, Switzerland}

\author{J. Kasparian}
\affiliation{Universit\'e de Gen\`eve, GAP-Non-linear, Chemin de Pinchat 22, CH-1211 Geneva 4, Switzerland}

\author{J.-P. Wolf}
\affiliation{Universit\'e de Gen\`eve, GAP-Biophotonics, Chemin de Pinchat 22, CH-1211 Geneva 4, Switzerland}

\date{\today}

\begin{abstract}
We show that multiple filamentation patterns in high-power laser beams, can be described by means of two statistical physics concepts, namely self-similarity of the patterns over two nested scales, and nearest-neighbor interactions of classical rotators. The resulting lattice spin model perfectly reproduces the evolution of intense laser pulses as simulated by the Non-Linear Schr\"odinger Equation, shedding a new light on multiple filamentation. As a side benefit, this approach drastically reduces the computing time by two orders of magnitude as compared to the standard simulation methods of laser filamentation.
\end{abstract}

\maketitle

The non-linear Schr\"odinger equation (NLSE), originally emanating from quantum mechanics, is paradigmatic of a universal equation which is widely used in a variety of fields such as non-linear optics~\cite{Gedalin1997Optical}, Bose-Einstein condensates~\cite{Ruprecht1995Time,Bradley1995Evidence}, plasma physics~\cite{Mio1976Modified}, or fluid mechanics~\cite{Dysthe1979Note}. Its analytical properties are quite well-known, and exhibit features such as integrability in one dimension~\cite{Chen1979Integrability}, or finite-time blow-up for higher spatial dimensions~\cite{Glassey1977Blowing,Fibich1999Self}.

In the field of non-linear optics, the NLSE describes light filaments \cite{ChinHLLTABKKS2005,Couairon2007Review} forming in the propagation of laser pulses which power exceeds a certain critical value. For powers much beyond the latter, the beam breaks up into many cells, each generating one filament \cite{CampiSS1973,M'ejKYSFWSVNCB2005,B'ejBEMWALSSKBCGBBBCCHLMMMPPRR2007}, forming complex multiple filamentation patterns \cite{HeninPKWJKBSSNSRWSMBS2010a}. We recently showed that the formation of such patterns from an initially smooth laser beam profile defines a two-dimensional phase transition governing the geometrical structuring of the beam and the self-organization of light filaments~\cite{Ettoumi2015Percofil}. The patterns associated to this phase transition are similar to those produced by percolation~\cite{Golden1998Percolation,Stauffer1979Scaling} or spin models from the statistical physics literature~\cite{Marro2005nonequilibrium,Kosterlitz1973Ordering,Rogers1989Cahn}. 

The salient features of such systems generally arise from the nearest-neighbor interactions between the underlying constituents, mainly quantum or classical spins. However, the description of multiple filamentation patterns as the result of basic interacting elements like spins was never considered until now. Laser filaments have been shown to laterally interact with their neighbors located at a distance of several millimeters  in the beam profile \cite{RenHFDM2000,BergAGSW2003,HosseLFLCKPAK2004,MaLXGZ2008,D'AsaHPAGSA2009,Berge1997Beamlets}. This interaction is attractive if the filaments are in phase, and repulsive if they are in antiphase~\cite{XiLZ2006,ShimSHVHIG2010}, because it is mediated by interference of the photon bath surrounding each filament~\cite{LiuTAGBC2005,CourvBKSMYW2003,KolesM2004,SkupiBPL2004}. However, such interactions have up to now been only considered locally. No impact on the global beam profile evolution was investigated, or even expected.

In this Letter, we derive a model for laser multiple filamentation, showing that this physical phenomenon can be understood as a consequence of self-similarity and nearest-neighbor interaction between coarse-grained light elements. This results in a description highly reminiscent of the Edwards-Anderson spin-glass model~\cite{Edwards1975EA,Almeida1978StabilityEA,Cugliandolo1993AnalyticalEA,Cugliandolo1994SK}, quantitatively bridging non-linear optics to out-of-equilibrium statistical physics.

In the following, we will first discuss the self-similarity of multiple filamentation patterns. Then, we will show that it allows to drastically coarse-grain the dynamics with minimal loss of information, provided time is adequately rescaled to account for the change in the speed of transverse information flow induced by this procedure. The resulting lattice spin model will then be validated by a direct confrontation to the results of the standard NLSE integration, showing an amazing agreement.

The starting point of our derivation is the NLSE, which, in dimensionless units, reads
\begin{equation}
\mathrm{i} \partial_\eta \psi + \Delta \psi + f(|\psi|^2)\psi= 0,
\label{eqn:NLSE_Full}
\end{equation}
where $\eta$ is the propagation distance, $\Delta \equiv \partial^2_x + \partial^2_y$ the two-dimensional transverse Laplacian which accounts for geometrical diffraction, and the function~$f$ describes the nonlinear physical mechanisms at play, including dissipation and saturation. Although the NLSE is ubiquituous in physics, in the following we mainly focus on the case of multiple filamentation, where pattern formation is best characterized both experimentally~\cite{Henin2010a} and theoretically~\cite{Ettoumi2015Percofil}. In filamentation, $\psi$ is the electric field envelope, and for numerical simulations, it is quite common to model the non-linearity as
\begin{equation}
f(|\psi|^2)=|\psi|^2 -|\psi|^{2K}+\mathrm{i}\nu |\psi|^{2K-2},
\label{eqn:f_phys}
\end{equation}
where the first term accounts for the Kerr self-focusing effect, and the two last ones model defocusing by free electrons as well as losses due to the $K$-photon ionization releasing these electrons. Without these last two terms, some initial conditions of Eq.~(\ref{eqn:NLSE_Full}) exhibit finite-time divergence~\cite{Fibich2001NLSBlowUp}.

Equation~(\ref{eqn:NLSE_Full}) features a linear instability, called the modulational instability, with spectacular experimental consequences ranging from the emergence of solitons in Bose-Einstein Condensates~\cite{Carr2004Spontaneous,Denschlag2000Generating} to the formation of multiple filamentation patterns~\cite{Kandidov1999,Moloney1999Turbulence,Hosseini2004,Ren2000} in large high-power laser beams. The growth rate~$\gamma$ of this instability can be obtained analytically. For a plane wave steady-state~$\psi_0 \mathrm{e}^{\mathrm{i}\lambda \eta}$, writing~$k_\perp$ the spatial transverse wave-vector of the perturbation leads to~\cite{BergAGSW2003}
\begin{equation}
\gamma = k_\perp\sqrt{2\psi_0^2 f^\prime(\psi_0^2)-k_\perp^2}.
\label{eqn:gamma_full}
\end{equation}

\begin{figure}[htbp]
	\centering
		\includegraphics[width=1.0\columnwidth]{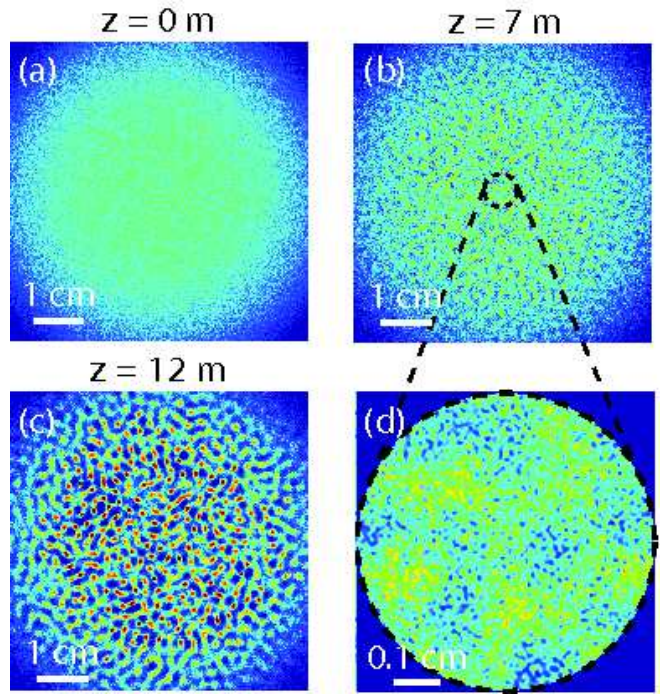} \\				
	\caption{(Color online) (a), (b), and (c) Evolution of an initially perturbed fourth order super-gaussian (flat top) laser profile of $50$~TW at $800$~nm, with $5$~cm waist. The modulational instability seeds the emergence of a pattern, which self-sustains when non-linear effects come into action. (d) Magnification of a central zone showing self-similarity.}
	\label{fig:evo_pattern}
\end{figure}

Figure~\ref{fig:evo_pattern} displays the resulting patterns in the case of laser propagation in air by solving numerically Eq.~(\ref{eqn:NLSE_Full}). The initial condition is taken as a fourth-order super-Gaussian of 5~cm diameter, holding $50$~TW at a wavelength of $800$~nm. The relationship between the dimensionless units and the real physical parameters is given in the supplementary information~\footnotemark[1]. The modulational instability, seeded by the initial beam noise (Fig.~\ref{fig:evo_pattern}a), triggers the emergence of mesoscopic structures (Fig.~\ref{fig:evo_pattern}b) which are later amplified (Fig.~\ref{fig:evo_pattern}c) by the non-linearities in Eq.~(\ref{eqn:f_phys}). Furthermore, Fig.~\ref{fig:evo_pattern}d displays a close-up of the center of the beam after 7~m of propagation. The patterns are quite similar at both scales. In particular, they share the following common features: (i) local maxima attracting intensity, depleting the energy around them; (ii) strings of intermediate intensity connecting these local maxima; (iii) regions of weaker intensity (photon bath) around them; (iv) lateral interactions between the maxima structures, and (v) the overall shrinking of the whole pattern towards a structure with the lower length scale. Similar patterns can therefore be observed on two spatial scales, two orders of magnitude apart in size.

\begin{figure}[htbp]
	\centering
		\includegraphics[width=1.0\columnwidth]{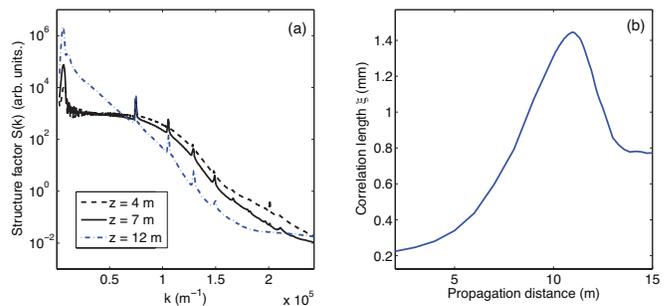}
	\caption{(Color online) (a) Structure factor~$S_\phi(k)$ calculated after $4$~m (representative of the initial conditions), $7$~m, and $12$~m of propagation. Note that we have suppressed the zero peak for clarity reasons. (b) Evolution of the correlation length~$\xi$ (Eq.~(\ref{eqn:xi})) for a $800~\mathrm{nm}$, $50~\mathrm{TW}$ beam of $5~\mathrm{cm}$ waist.}
	\label{fig:Evo_xi_spectra}
\end{figure}

Beyond the visual aspect, the self-similarity can be quantitatively evidenced by investigating the structure factor~$S_\phi$ of the laser fluence $A\equiv|\psi|^2$, defined by:
\begin{equation}
S_\phi(\mathbf{k},\eta)=\langle |\hat{A}(\mathbf{k},\eta)|^2 \rangle,
\end{equation}
where the hat symbol denotes the Fourier transform, and the brackets an ensemble average. Figure~\ref{fig:Evo_xi_spectra}a displays three spectra corresponding to two stages of the evolution of the laser beam. Starting from a flat transverse spectrum describing the various lengthscales of the initial profile modulated with a white noise, the modulational instability seeds the emergence of the characteristic patterns at stake here. The peaks on the spectrums after $7$ and $12$~m propagation depict the aforementioned multiple scales constitutive of the self-similarity.

Let us define the characteristic length~$\xi$ of the transverse patterns using the structure factor by the circular average
\begin{equation}
\xi(\eta) = \frac{\int S_\phi(k,\eta) \mathrm{d}k}{\int k S_\phi(k,\eta) \mathrm{d}k}.
\label{eqn:xi}
\end{equation}
During the propagation, $\xi$ first increases from the initial noise correlation length until a maximum length attained at the percolation threshold~\cite{Ettoumi2015Percofil} (Figure~\ref{fig:Evo_xi_spectra}b). This increase differs from the monotonic decay of the correlation length that is obtained with thresholded, two-color images~\cite{Ettoumi2015Percofil}. At further propagation distances, the fluence clusters either vanish because of dissipation, or get squeezed in size because of the energy flux towards their center, resulting in a decrease of~$\xi$. 

The images shown on Figure~\ref{fig:evo_pattern} are reminiscent of many models studied by the statistical mechanics community. For example, one can note a striking resemblance with coarsening phenomena~\cite{Bray1994Theory,Bray1994Coarsening}. In our case, the transverse low and high intensity regions can be seen as two different phases of a generic model evolving under Ginzburg-Landau dynamics~\cite{Berthier1999Response}.

Such a behavior is generally well reproduced by simple spin models with proper time dynamics, and we shall now derive lattice spin model (LSM) of filamentation, based on the previous observations. The patterns topology we aim at reproducing is mainly due to the combined effects of modulational instability and Kerr non-linearity. We therefore truncate $f$ (Eq.~(\ref{eqn:f_phys})) to its cubic contribution in~$\psi$.

The typical patterns shown in Figure~\ref{fig:evo_pattern} strongly suggest to model the electric field by a superposition of narrow, Gaussian-like, elementary wavelets. We chose their spatial extensions comparable to the lowest-order structure in the beam, i.e.~10~$\mu$m. Therefore, the field can be expanded as $\psi(\mathbf{r},\eta) = \sum_{n} A_n(\mathbf{r},\eta) \mathrm{e}^{\mathrm{i}\phi_n(\mathbf{r},\eta)}$.

The universality class unveiled in~\cite{Ettoumi2015Percofil} suggests that the behavior of a lattice model close to criticality should be independent from the microscopic detail, and \textit{a fortiori} from the lattice geometry. Hence, we define the set of $\lbrace\mathbf{r_n}\rbrace$ as a square lattice.
Projecting Eq.~(\ref{eqn:NLSE_Full}) on each $\mathrm{e}^{\mathrm{i}\phi_n}$ and identifying real and imaginary parts leads to
\begin{eqnarray}
\partial_\eta A_n &=& -2 \boldsymbol{\nabla}A_n \cdot \boldsymbol{\nabla}\phi_n- A_n \Delta \phi_n,\label{eqn:system11}\\
A_n\partial_\eta\phi_n &=& \Delta A_n - A_n \left|\boldsymbol{\nabla} \phi_n\right|^2 \label{eqn:system12}\\
 & & + A_n \sum_{\ell,m} A_\ell A_m \cos\left(\phi_\ell-\phi_m\right).\nonumber
\end{eqnarray}

By definition, $A_n$ displays a maximum at $\mathbf{r}=\mathbf{r_n}$. Considering that the phase~$\phi_n$ is strongly impacted by the $B$-integral~\cite{Couairon2007Review}, hence by the amplitude $A_n$, we assume that it also has an extremum at the same location. Therefore, Eqs.~(\ref{eqn:system11})-(\ref{eqn:system12}) can be simplified by cancelling every first spatial derivative. As detailed in the Supplementary Information~\footnote{\label{1}See Supplementary Material [url] for details about the lattice spin model derivation and time renormalization procedure.}, the spatial self-similarity allows us to define the each square lattice site~$n$ of area~$\delta^2$ as holding two observables, $A_n$ and $\phi_n$, defined as the respective averages of the amplitude and phase of the underlying small-scale wavelets of the considered cell:
\begin{eqnarray}
\dot{A}_n &=& - \kappa\left[\phi_n\right] A_n, \label{eqn:lattice_sys_1}\\
\dot{\phi}_n &=& \frac{\kappa\left[A_n\right]}{A_n} + A_n^2 + A_n \sum_{\langle\ell\rangle_n} A_\ell \cos\left(\phi_n-\phi_\ell\right), \label{eqn:lattice_sys_2}
\end{eqnarray}
where the dots in~(\ref{eqn:lattice_sys_1})-(\ref{eqn:lattice_sys_2}) refer to a ``time'' derivative, which will reproduce the propagation dynamics of the original NLSE, and where $\kappa$ is the discretized Laplacian over the four nearest-neighbors. For a site $(i,j)$, it reads $\kappa\left[\phi_{i,j}\right]=\frac{1}{\delta^2}\left(-4 \phi_{i,j}+\phi_{i+1,j}+\phi_{i-1,j}+\phi_{i,j+1}+\phi_{i,j-1}\right)$.

Eqs.~(\ref{eqn:lattice_sys_1})-(\ref{eqn:lattice_sys_2}) are the main result of this Letter. Each lattice site can be seen as an individual classical rotator, described by two observables~$A$ and $\phi$, which are its length and angle, respectively. These rotators evolve under nearest-neighbor interactions, arising from both the discretized Laplacians and the last term of Eq.~(\ref{eqn:lattice_sys_2}), accounting for the coarse-grained interference phenomenon. For instance, if two lattice neighbors share a common optical phase, their amplitudes will constructively interfere, mimicking the situation in which two filaments attract each other, and eventually merge. Conversely, if these two neighbors feature a relative phase shift of $\pi$, a destructive interference will decrease their amplitudes and eject their energy towards sites further away, mimicking the experimentally observed repulsion~\cite{ShimSHVHIG2010}.
 
The interaction term $J_{n\ell} \equiv \cos(\phi_n-\phi_\ell)$ is typical of the spin-glass model, e.g. the soft-spin version of the Edwards-Anderson model~\cite{Edwards1975EA,Almeida1978StabilityEA,Cugliandolo1993AnalyticalEA,Sompolinsky1982Relaxational}, characterized by the interaction Hamiltonian between spins~$\sigma_i$
$\mathcal{H} = - \sum_{\langle i,j \rangle} J_{ij} \sigma_i \sigma_j$ and evolving under a phenomenological Langevin equation such as
\begin{equation}
\dot{\sigma_i} = -\beta \dfrac{\delta \mathcal{H}}{\delta \sigma_i} + \xi_i = \beta \sum_{\langle j \rangle_i} J_{ij} \sigma_j + \xi_i,
\end{equation}
$\beta$ being the inverse temperature and $\xi_i$ a Gaussian random variable.

As a test for the relevance of the presented Lattice Spin Model (LSM), we will now compare its pattern predictions with the results obtained by integrating the NLSE using a standard Split-Step Fourier Method (SSFM). As an initial condition, we will use an already slightly propagated beam (by $4$~m) with the same properties as in Figure~\ref{fig:evo_pattern}. In general, one could estimate the coarse-graining length~$\delta$ as being the inverse of the wavelength maximizing the linear growth rate~$\gamma$, since clusters of such a size are expected to emerge quicker than others. Doing so with an initial condition as presented here yields $\delta=924~\mu$m, in good agreement with the actual physical range as the size of the photon bath surrounding a single filament is typically between 500 and 1000~$\mu$m. Practically speaking, the coarse-graining leading to the definition of the spins ($A_n,\phi_n$) from the field~$\psi$ writes as
\begin{eqnarray}
A_n &=& \left(\frac{1}{\delta^2} \iint_{\Sigma_n} |\psi(\mathbf{r_n+r^\prime})|^2 \mathrm{d}^2\mathbf{r^\prime}\right)^{1/2},
\label{eqn:CG_procedure} \\
\phi_n &=& \frac{1}{\delta^2} \iint_{\Sigma_n} \phi(\mathbf{r_n+r^\prime}) \mathrm{d}^2\mathbf{r^\prime},
\label{eqn:CG_procedure_phi}
\end{eqnarray}
where $\Sigma_n$ stands for the lattice cell~$n$ of area~$\delta^2$. Note that it is important to average the fluence~$|\psi|^2$, and then only take the square root instead of directly averaging the field~ $\psi$. This way, Eq.~(\ref{eqn:CG_procedure}) ensures the conservation of the photon number~$P_0 = \sum_{n} \delta^2 A_n^2$.

A direct integration of Eqs.~(\ref{eqn:lattice_sys_1})-(\ref{eqn:lattice_sys_2}) yields a very good qualitative agreement with the reference NLSE patterns computed with the SSFM. However, the smallest coarse-grained length scales of order~$\delta$ (typically millimetric) behave within the same timescale in the LSM as their much smaller counterparts of a hundred micrometers in the NLSE hereafter denoted by~$\ell_\mathrm{c}$. As a consequence, a pattern arising after a few meters would be predicted after only a few centimeters by the LSM.

However, a linear stability analysis shows that the LSM exhibits the same growth rate given by~Eq.(\ref{eqn:gamma_full}) as the NLSE, which is remarkable. Based on this result, we devised a strategy detailed in the supplementary information~\footnotemark[1] in order to recover the proper dynamics, introducing a rescaled ``time'' variable~$\tau$ reading
\begin{equation}
\tau = \eta \sqrt{\frac{\ell_\mathrm{c}}{\delta}}.
\label{eqn:renorm_time}
\end{equation}
This time renormalization therefore ensures that the LSM correctly reproduces the speed of the transverse diffusion of information.

In our case, we considered a coarse-graining length $\delta=732~\mu\mathrm{m}$, i.e. $40$ pixels in our reference NLSE numerical resolution. Since we smoothed our initial random noise over a length of~$4$ pixels, we set the small-scale cutoff length $\ell_\mathrm{c}=\delta/10=73.2~\mu\mathrm{m}$. From Eq.~(\ref{eqn:renorm_time}), we deduce that the time renormalization factor is equal to~$0.32$. Again, this factor lower than unity translates the fact that the coarse-graining causes the LSM to act on the patterns much quicker than the NLSE does, since the former is an upscaled version of the latter.

\begin{figure}
	\centering
		\includegraphics[width=1.0\columnwidth]{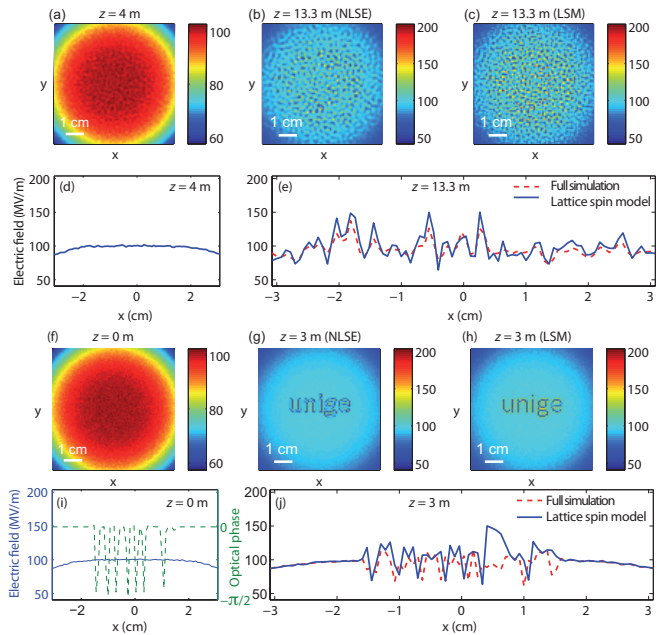}
	\caption{(Color online) Comparison between the lattice spin model and the coarse-grained result of the NLSE simulation using the SSFM. (a-e), flat initial phase; (f-j), worst-case scenario with $\pi / 2$ phase jumps. (a), (f) Initial condition for the LSM, originating from a full resolution SSFM integration; (b), (g) LSM output; (c), (h) NLSE output using SSFM; (e), (j) Horizontal cuts across the model ouptuts.}
	\label{fig:Comparaison_Amplitude_phase}
\end{figure}

We first assess the validity of the LSM by considering a flat initial phase, namely a real initial~$\psi$. Figure~\ref{fig:Comparaison_Amplitude_phase} (a-e) compares the fluence pattern obtained from both the LSM and the NLSE after approximately $9$~m of free propagation. Despite the apparent lack of information in the initial condition (at~$4$~m), the LSM remarkably reproduces the final reference NLSE pattern, showing that the interpretation of multiple filamentation in terms of interacting spins yields quantitative predictions.This is very remarkable as filamentation is generally considered as a local phenomenon requiring a high spatial resolution in order to capture its salient features.

To check the fidelity of the phase evolution in a ``worst-case scenario'', we also considered initial abrupt phase jumps from zero to $-\pi/2$ (Figure~\ref{fig:Comparaison_Amplitude_phase}i). This case was chosen because of the difficulty to simulate fields with steep gradients, which result in strong diffraction and instabilities that can only be resolved at extremely high resolutions. As a coarse-grained model intrinsically cannot capture such features, such a situation should check the robustness of our lattice spin description.

We modulated the aforementioned amplitude pattern by a phase mask displaying the word ``unige''. Figure~\ref{fig:Comparaison_Amplitude_phase} (g), (h), and (j) shows a remarkable agreement after 3~m of free propagation, despite a glitch on the reproduction of the letter ``g''.

Moreover, these results are not tributary to a fine-tuned choice of the coarse-graining length~$\delta$. One can freely choose it in the aforementioned physically acceptable range and still obtain reasonable results, whereas a decrease of resolution in the SSFM method rapidly leads to erroneous simulations.

These two test cases highlight the relevance and even the quantitative accuracy of the LSM. For smooth initial phases, the relative error on intensity stays below 10\%. The importance of nearest-neighbor interaction was further demonstrated by switching off the corresponding term in Equation~(\ref{eqn:lattice_sys_2}). The beam then keeps a smooth shape very different from the self-structuring of the beam observed in both experiments and NLSE simulations.

It is quite straightforward to derive richer lattice models encompassing more phenomena, such as e.g. other non-linearities, saturation mechanisms, plasma generation (see Eq.~(\ref{eqn:f_phys})) or even air turbulence. This would simply require to expand the additional physical model on the wavelet basis, and then simplify all the remaining terms by keeping in mind the nearest-neighbor picture.

As a conclusion, we took advantage of the self-similarity of multiple filamentation patterns to introduce the description of laser multiple filamentation as a Lattice Spin Model with glassy-like dynamics. The numerical benchmarks showed an excellent agreement with the full calculations, demonstrating the robustness of such a novel interpretation, that can also be related to the recent observation of a percolation-like phase transition in such a system~\cite{Ettoumi2015Percofil}. Furthermore, as a consequence of the coarse-grained description, the small lattice sizes at play allow computing times faster by two orders of magnitude as compared to standard SSFM calculations. Such a speed-up opens the way to statistical studies of, e.g., beam propagation through turbulence, or explicit inversion of non-linear Lidar measurements~\cite{Hemmer2011,NatanLGKS2012,Bremer2013} of atmospheric trace constituents.

In a wider scope, our approach only relies on the structure of the NLSE, not on a particular nonlinearity (i.e. a particular function~$f$), nor its application to a specific physical system. Therefore, it can be generalized to other fields of physics described by the NLSE, where such self-similarity could also be observed and exploited~\cite{Bao2003Numerical,Fort2005BECRandom,Modugno2006BECRandom}.

\begin{acknowledgments}
We wish to warmly thank an anonymous referee for very valuable comments and suggestions. We gratefully acknowledge fruitful discussions with T.~Giamarchi, M.~Brunetti and S.~Hermelin. We acknowledge financial support from the European Research Council Advanced Grant ``Filatmo'' and the Swiss National Science Foundation (Grant 200021-155970).
\end{acknowledgments}

\section{Dimensionless units}

Considering the non-linearity function~$f(|\psi|^2)=|\psi|^2 -|\psi|^{2K}+\mathrm{i}\nu |\psi|^{2K-2}$,
where $K$ is the number of photons required for medium ionization, the dimensionless units read
\begin{eqnarray}
\eta&=&z(\gamma/(\alpha k_0 n_2)^K)^{-1/(K-1)},\\
\psi&=&A (\alpha k_0 n_2/\gamma)^{-1/(2K-2)},\\
\tilde{x}&=&x\sqrt{2k_0} \left(\gamma/(\alpha k_0 n_2)^K\right)^{-1/(2K-2)},
\end{eqnarray}
where $x$ (transverse coordinate) and $z$ (propagation distance) are in~meters, $A$ (the electric field envelope) in $\mathrm{V}.\mathrm{m}^{-1}$. $k_0=2\pi/\lambda_0$ is the central wavenumber in $\mathrm{m}^{-1}$, for which we use $\lambda_0=800$~nm. The non-linear refractive index $n_2$ is set to $1.2\times10^{-23}~\mathrm{m}^2.\mathrm{W}^{-1}$. The parameters $\alpha$ and $\gamma$ describe the medium delayed response, and can be found in reference~\cite{Skupin2004Multifil}.

\section{Derivation of the lattice model equations}

\begin{figure}[htbp]
	\centering
		\includegraphics[width=1.0\columnwidth]{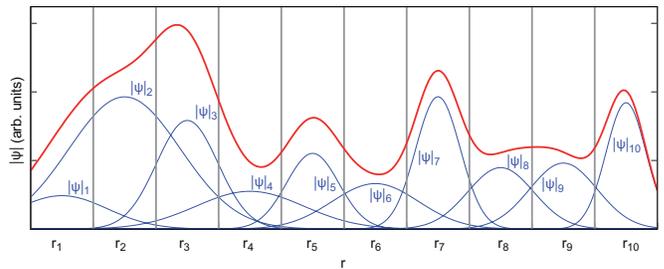}
	\caption{Principle of the wavelet decomposition and lattice discretization procedure. Each wavelet~$k$ of the decomposition is supposed to be centered on $\mathbf{r_k}$.}
	\label{fig:wavelets_principe}
\end{figure}

In this section, we shall explicit the derivation of the lattice model starting from the observation that the electric field~$\psi$ can be developed on a basis of Gaussian wavelets, as emphasized in Figure~\ref{fig:wavelets_principe}.
Practically speaking, we expand the field as
\begin{equation}
\psi(\mathbf{r},\eta) = \sum_{n} A_n(\mathbf{r},\eta) \mathrm{e}^{\mathrm{i}\phi_n(\mathbf{r},\eta)}.
\label{eqn:wavelet}
\end{equation}
Injecting this decomposition into the NLSE, one obtains the following system:
\begin{eqnarray}
\partial_\eta A_n &=& -2 \boldsymbol{\nabla}A_n \cdot \boldsymbol{\nabla}\phi_n- A_n \Delta \phi_n,\label{eqn:system11_SM}\\
A_n\partial_\eta\phi_n &=& \Delta A_n - A_n \left|\boldsymbol{\nabla} \phi_n\right|^2 \\
& & + A_n \sum_{\ell,m} A_\ell A_m \cos\left(\phi_\ell-\phi_m\right).\nonumber
\label{eqn:system12_SM}
\end{eqnarray}

The extrema of $A_n$ and $\phi_n$ at each beam center $\mathbf{r}=\mathbf{r_n}$ allow to cancel out every first spatial derivative at these locations. The system~(\ref{eqn:system11_SM})-(\ref{eqn:system12_SM}) then rewrites as:
\begin{eqnarray}
\partial_\eta A_n &=& - \left(\Delta \phi_n\right) A_n,\label{eqn:system21_SM}\\
\partial_\eta \phi_n &=& \frac{\Delta A_n}{A_n} +  \sum_{\ell,m} A_\ell(\mathbf{r_n}) A_m(\mathbf{r_n}) \label{eqn:system22_SM}\\
& &\times\cos\left[\phi_\ell(\mathbf{r_n})-\phi_m(\mathbf{r_n})\right].\nonumber
\end{eqnarray}
As the width of the individual wavelets has been chosen comparable with that of the lattice cells, interactions between elementary beams can be neglected except for the four nearest neighbors, greatly simplifying the last term in Eq.~(\ref{eqn:system22_SM}):
\begin{eqnarray}
\partial_\eta \phi_n &=& \frac{\Delta A_n}{A_n} + A_n^2 \label{eqn:phase2_SM}\\
& &+ A_n \sum_{\langle\ell\rangle_n} A_\ell(\mathbf{r_n}) \cos\left[\phi_n-\phi_\ell(\mathbf{r_n})\right]\nonumber,
\end{eqnarray}
where the notation~$\langle \ell \rangle_n$ denotes a summation over the nearest-neighbors on the square lattice of the $n^{\text{th}}$ elementary beam. Note that the second term in the right-hand side of Eq.~(\ref{eqn:phase2_SM}) corresponds to the standard $n_2 I$ cumulative phase shift term, also known as the $B$-integral. This phase evolution equation is similar to a short-range disordered Kuramoto model~\cite{Kuramoto1987SRKura,Acebron2005KuramotoReview}, a paradigmatic model for synchronization.

We shall now take benefit of the self-similarity of our problem to upscale the model from the $100~\mu$m microscopic filamentary structures, to that of the millimeter sized aggregates. In that purpose, we define a new lattice, with coarser cells of width hereafter denoted by $\delta$.
To get rid of the continuous spatial representation~$\mathbf{r}$ in Eqs.~(\ref{eqn:system21_SM}) and~(\ref{eqn:phase2_SM}), we replace the Laplacian operator $\Delta$ by its square-lattice discretized counterpart, named~$\kappa$, which action on a site $n\equiv(i,j)$ reads
\begin{equation}
\kappa\left[\phi_{i,j}\right]=\frac{1}{\delta^2}\left(-4 \phi_{i,j}+\phi_{i+1,j}+\phi_{i-1,j}+\phi_{i,j+1}+\phi_{i,j-1}\right).
\end{equation}
We also replace the wavelets defined by $A_\ell(\mathbf{r_n})$ (resp. $\phi_\ell(\mathbf{r_n})$) by $A_\ell(\mathbf{r_\ell})$ (resp. $\phi_\ell(\mathbf{r_\ell})$. Note that this is a strong assumption, but essential for the final nearest-neighbor interacting model, which finally reads
\begin{eqnarray}
\dot{A}_n &=& - \kappa\left[\phi_n\right] A_n, \label{eqn:lattice_sys_1_SM}\\
\dot{\phi}_n &=& \frac{\kappa\left[A_n\right]}{A_n} + A_n^2 + A_n \sum_{\langle\ell\rangle_n} A_\ell \cos\left(\phi_n-\phi_\ell\right). \label{eqn:lattice_sys_2_SM}
\end{eqnarray}

\section{Time renormalization}

In this section, we shall demonstrate the correct time rescaling so as to obtain the correct propagation dynamics with the lattice model.
Neglecting the last nearest-neighbour coupling in equation~(\ref{eqn:lattice_sys_2}), we find that like for the NLSE, the plane-wave defined by the homogeneous amplitude $A=A^\ast$ and phase $\phi^\ast=\phi+{A^\ast}^2 t$ is a steady-state solution. Furthermore, linearizing the system~(\ref{eqn:lattice_sys_1_SM})-(\ref{eqn:lattice_sys_2_SM}) and considering the first-order corrections of $\delta A$ and $\delta \phi$ yields, in the Fourier space,
\begin{equation}
\dfrac{\partial}{\partial t}
\left(
\begin{array}{c}
\delta \hat{A} \\
\delta \hat{\phi}
\end{array}\right) = 
\left(
\begin{array}{c c}
0 & k^2 A^\ast \\
2A^\ast-\frac{k^2}{A^\ast} & 0
\end{array}\right)
\left(
\begin{array}{c}
\delta \hat{A} \\
\delta \hat{\phi}
\end{array}\right)
\end{equation}
The eigenvalues of the Jacobian matrix are given by
\begin{equation}
\gamma=\pm k\sqrt{2{A^\ast}^2-k^2},
\label{eqn:gamma_SM}
\end{equation}

In Fourier space, the spatial averaging acts as a low-pass filter, and brutally cuts all spatial frequencies above $k_\mathrm{c}=2\pi/\delta$. As a result, the growth rates calculated for both the original pattern and its coarse-grained counterpart might differ because of the change in the spectrum distribution, leading to an erroneous description of the dynamics by the lattice model. In order to tackle this issue, we rescaled the time variable in the lattice model so as to recover the same apparent growth as for the full equation.

More specifically, we rescaled time in the coarse-grained case so that the Fourier-averaged initial time derivatives of the growth rate coincide for both cases. Namely, we look for a time variable~$\tau$ satisfying
\begin{equation}
\dfrac{\partial}{\partial \eta} \left[ \int \hat{\psi}(k,\eta) \mathrm{d}k \right] \Big|_{\eta=0} = \dfrac{\partial}{\partial \tau} \left[ \int \hat{A}(k,\tau) \mathrm{d}k \right] \Big|_{\eta=0}.
\label{eqn:CI_renorm}
\end{equation}
This approach is motivated by the fact that we want the lattice model to reproduce the dynamics of the NLSE already in the beginning of the propagation. For instance, we chose to rely more on the time derivative at the origin rather than the maximum growth rate.

Plugging the fields' expressions in the linear regime in Eq.~(\ref{eqn:CI_renorm}), we obtain
\begin{equation}
\int \hat{\psi}(k,\eta=0) \gamma(k) \mathrm{d}k = \int \hat{A}(k,\tau=0) \gamma(k) \mathrm{d}k,
\end{equation}
where the growth rate~$\gamma$ is given in each model by Eq.~(\ref{eqn:gamma_SM}).

In order to express~$\tau$ with respect to~$\eta$, let us consider a full resolution initial condition $\psi_0$ exhibiting a flat spectrum, with a high-frequency cutoff~$2\pi/\ell_\mathrm{c}$. In this case, the structure factor reads $S_\phi=S_0 \chi_{[0,2\pi/\ell_\mathrm{c}]}$, where $\chi_E$ is the indicatrix of the set~$E$. This corresponds to a plane wave perturbed by a white noise with a correlation length of~$\ell_\mathrm{c}$. In virtue of Parseval's theorem, the spectral power amplitude relates to the initial power~$P_0$ by
\begin{equation}
P_0 = \frac{2 \pi}{\ell_\mathrm{c}} S_0,
\end{equation}
After the coarse-graining procedure, still in virtue of Parseval's theorem, we have
\begin{equation}
\sum_{k} \delta^2 |\hat{A}_k|^2 = P_0,
\label{eqn:Parseval_2_SM}
\end{equation}
where the summation is performed over the reciprocal lattice, $\hat{A}$ being the Fourier-transformed coarse-grained field. Since the frequencies higher than $2\pi/\delta$ have been wiped out from the new spectrum by the coarse-graining, Eq.~(\ref{eqn:Parseval_2_SM}) implies that the new spectral power amplitude~$S^{\mathrm{cg}}_0$ must read
\begin{equation}
S^{\mathrm{cg}}_0=S_0 \frac{\delta}{\ell_\mathrm{c}}.
\end{equation}
Since $\delta>\ell_\mathrm{c}$, the new spectrum displays a higher amplitude in order to ensure the photon number conservation, despite the disappearance of high-frequency modes, that corresponded to the microscopic detail that we got rid of by spatial averaging.

It is now our aim to calculate the average growth rate~$\langle \gamma \rangle$ for both cases, given the two different spectral probability distributions
\begin{eqnarray}
f(k) &=& \sqrt{\frac{\ell_\mathrm{c}}{2\pi P_0}} \chi_{[0,2\pi/\ell_\mathrm{c}]},\label{eqn:Pk_full_SM} \\
f^{\mathrm{cg}}(k) &=& \sqrt{\frac{\delta}{2\pi P_0}} \chi_{[0,2\pi/\delta]}.\label{eqn:Pk_cg_SM}
\end{eqnarray}
Let us calculate the probability distribution of the restriction of~$\gamma$ to its real positives values. The probability of observing a growth rate lower than $\gamma$ for a random $k$ chosen with probability~$f$ (which can be given either by Eqs.~(\ref{eqn:Pk_full_SM}) or~(\ref{eqn:Pk_cg_SM})) then reads
\begin{equation}
P(y\leq \gamma) = \int_0^{\sqrt{A^2-\sqrt{A^4-\gamma^2}}} f(k) \mathrm{d}k + \int_{\sqrt{A^2+\sqrt{A^4-\gamma^2}}}^{A\sqrt{2}} f(k) \mathrm{d}k.
\end{equation}
After differentiation with respect to $\gamma$, one obtains the probability distribution function for the positive growth rate. It is then straightforward to understand that the coarse-grained average growth rate relates to its original counterpart through
\begin{equation}
\left\langle \gamma^{\mathrm{cg}} \right\rangle = \left\langle \gamma \right\rangle \sqrt{\frac{\delta}{\ell_\mathrm{c}}},
\end{equation}
which gives the final relationship between the time variable~$\tau$ in the lattice model and the original distance variable~$\eta$ in the NLSE,
\begin{equation}
\tau = \eta \sqrt{\frac{\ell_\mathrm{c}}{\delta}}.
\label{eqn:renorm_time_SM}
\end{equation}

\section{Computational considerations}

Given a $N\times N$ square lattice, the model integration is of complexity $O(N^2)$, while the SSFM, helped by the Fast-Fourier-Transform algorithms available at hand, has a complexity scaling as $O(N^2\mathrm{log} N)$, which is slightly greater.

The result for $9$ meters of propagation for a realistic field as presented in the main article is obtained after only 35~s of computational time on a desktop computer, while the NLSE took 5~h on a dedicated workstation, without GPU acceleration.

Even if the recent GPU accelerators can greatly improve the computational performance of the SSFM, the resolution and complexity of the lattice model are far too low to compare. The lattice model provides a new paradigm for laser multiple filamentation, the computational speed-up is a consequence of such a simplification, though at the expense of the microscopic detail.

\bibliography{KuraBib}

\end{document}